\documentclass[aps,english,prl,reprint,longbibliography,notitlepage,superscriptaddress,preprintnumbers,twocolumn,showkeys]{revtex4-2}

\usepackage{graphicx}
\usepackage{bm}
\usepackage{hyperref}
\usepackage{amsmath}
\usepackage{amssymb}
\usepackage{microtype}
\usepackage{xcolor}
\usepackage[capitalise]{cleveref}
\usepackage{bbm}
\usepackage{mathtools}
\usepackage{ulem}
\usepackage{dutchcal}
\usepackage{tensor}

\newcommand{\dd}{\mathrm{d}}

\DeclareMathAlphabet{\mathpzc}{OT1}{pzc}{m}{it}

\newcommand{\para}[1]{\par\vspace{2mm}\noindent\textbf{#1}\,---\,}

\newcommand{\parait}[1]{\par\vspace{2mm}\noindent\textit{#1}\,---\,}

\makeatletter
\DeclareRobustCommand{\rcite}[1]{%
  \rcite@aux#1,\@nil{#1}%
}
\def\rcite@aux#1,#2\@nil#3{%
  \if\relax#2\relax
    Ref.~\cite{#3}%
  \else
    Refs.~\cite{#3}%
  \fi
}
\makeatother

\hypersetup{
    colorlinks = true,
    citecolor = {red},
    linkcolor = {blue},
    urlcolor = {blue},
}

\begin{document}

\title{
To CPL, or not to CPL? What we have not learned about the dark energy equation of state}

\author{Savvas Nesseris}
\email{savvas.nesseris@csic.es}
\affiliation{Instituto de F\'isica Te\'orica (IFT) UAM-CSIC, C/ Nicol\'as Cabrera 13-15, Campus de Cantoblanco UAM, 28049 Madrid, Spain}

\author{Yashar Akrami}
\email{yashar.akrami@csic.es}
\affiliation{Instituto de F\'isica Te\'orica (IFT) UAM-CSIC, C/ Nicol\'as Cabrera 13-15, Campus de Cantoblanco UAM, 28049 Madrid, Spain}
\affiliation{CERCA/ISO, Department of Physics, Case Western Reserve University, 10900 Euclid Avenue, Cleveland, OH 44106, USA}
\affiliation{Astrophysics Group \& Imperial Centre for Inference and Cosmology, Department of Physics, Imperial College London, Blackett Laboratory, Prince Consort Road, London SW7 2AZ, United Kingdom}

\author{Glenn D. Starkman}
\email{glenn.starkman@case.edu}
\affiliation{CERCA/ISO, Department of Physics, Case Western Reserve University, 10900 Euclid Avenue, Cleveland, OH 44106, USA}

\date{\today}

\begin{abstract}
We show that using a Taylor expansion for the dark energy equation-of-state parameter and limiting it to the zeroth and first-order terms, i.e., the so-called Chevallier-Polarski-Linder (CPL) parametrization in regimes where it has been shown to fail as a physics-based two-parameter model, instead of allowing for higher-order terms and then marginalizing over them, adds extra information not present in the data and leads to markedly different and potentially misleading conclusions. Fixing the higher-order terms to zero, one concludes that vacuum energy that is currently non-dynamical (e.g., the cosmological constant) is excluded at several $\sigma$ significance as the explanation of cosmic acceleration, even in Dark Energy Spectroscopic Survey (DESI) DR1 data. Meanwhile, instead marginalizing over the higher-order terms shows that we know neither the current dark energy equation of state nor its current rate of change well enough to make  such a claim. The CPL parametrization also implies that dark energy exhibits phantom behavior at high redshifts, while we show that by allowing the higher-order terms---which is required in order to capture the behavior of the dark energy equation of state in regimes beyond the validity of the CPL parametrization---the evidence for this phantom-like dark energy significantly weakens. This is not an argument for the  higher-order phenomenological parametrizations, but rather a caution regarding such parametrizations in general.
This issue has become more prominent now with the recent release of high-quality Stage IV galaxy survey data. The results of analyses using simple parametrizations should be interpreted with great care. 
\end{abstract}


\preprint{IFT-UAM/CSIC-25-29}

\maketitle

\para{Introduction.}
Nearly three decades after the announced discovery of cosmic acceleration \cite{SupernovaCosmologyProject:1998vns,SupernovaSearchTeam:1998fmf}, most cosmologists harbor little doubt that this acceleration is real. 
There is much less agreement about the cause of that acceleration.
The explanation that sits squarely within general relativity (GR), namely the cosmological constant,  is parametrically the simplest possibility---it requires a single constant of nature.

Despite the parsimony of the cosmological constant, the extremely tiny value that the observed acceleration implies for the energy density of the quantum vacuum challenges many fundamental physicists' credulity.
No matter whether one measures that energy density in Planck units ($2\times10^{-123}M_\mathrm{Pl}^4$ in units where $\hbar=c=1$) or in units more connected to the scale of the Standard Model of particle physics, the  vacuum expectation value of the Higgs field ($7\times 10^{-57}v_\mathrm{EW}^4$), the cosmological constant is extraordinarily small. 
The origin of the minuscule ratio of the vacuum energy to any simple estimate of it, known as the cosmological constant problem \cite{Martin:2012bt,Burgess:2013ara}, has puzzled theoretical physicists since the inception of GR \cite{Nernst_vacuumenergy}. 
So great is the interest in this confounding question that a preferred anthropic perspective drives  \cite{Weinberg:1988cp} to be Steven Weinberg's second-most-cited paper.

The lack of a compelling explanation of the value of the cosmological constant 
has driven the development of a menagerie of alternatives that fall under the broad rubric of dark energy \cite{Copeland:2006wr}. 
More recently, the difficulty string theorists have encountered in identifying a string vacuum state with positive energy density has led to the swampland conjecture \cite{Obied:2018sgi} that this is not a failure of string theorists but a feature of string theory.
Whichever the motivation of their builders, these models all share the property that they predict that the effective vacuum energy density will be time-dependent, or equivalently that the equation-of-state parameter, 
the ratio of the effective pressure of the vacuum to the effective energy density of the vacuum,
\begin{equation}
    w_\mathrm{vac} = \frac{P_\mathrm {vac}}{\rho_\mathrm{vac}}\,,
\end{equation}
will not have the constant value $w=-1$ characteristic of a static Lorentz-invariant vacuum energy density.
(They also typically require an unspecified solution to the cosmological constant problem i.e., they depend on an unspecified explanation why the Standard Model contribution to the vacuum energy does not overwhelm by tens of orders of magnitude the contributions they introduce.)

Observation of the time dependence of the vacuum energy density would be a clear signal that the cosmic acceleration is not, or at least not solely, due to the cosmological constant.  
Cosmologists have taken up this challenge from fundamental physicists with gusto. 
Many of the current and planned galaxy surveys have as a primary objective to measure the redshift dependence of the dark energy equation-of-state parameter $w_\mathrm{DE}(z)$.
The precise observational handles on $w_\mathrm{DE}(z)$ vary---galaxy number counts, supernova luminosity distances, weak lensing measurements, etc.---but the target remains the nature of dark energy.

For any specific model of the dark energy, one can, hopefully, make specific predictions for $w_\mathrm{DE}(z)$ as a function of the model parameters, which are typically a mix of Lagrangian parameters and  the initial values of newly introduced fields. 
One can then compute the likelihoods for the observational data given the model,  obtain best-fit parameter values, and make information-criteria-based comparisons between the model and all other models, including a simple cosmological constant.
Unfortunately, we are not yet in a situation where there is consensus that we have identified  the full set of reasonable models.  
Many observational cosmologists would prefer, therefore, to present their results in a model-agnostic form and leave it to others to perform the subtle model comparisons necessary to identify the most compelling model.
Moreover, such model-agnostic quantities may be more satisfactory figures of merit by which to characterize an ambitious observational program to the community and to funding agencies.

In the context of $w_\mathrm{DE}(z)$, a model-agnostic form means some parametrization of this function in terms of a small set of numbers. 
Several such parametrization have been considered, but the one that is currently almost always used  to report the results of cosmological experiments is the Chevallier-Polarski-Linder (CPL) parametrization \cite{Chevallier:2000qy,Linder:2002et}.
In terms of the scale factor $a$ of the Friedmann-Lema\^{i}tre-Robertson-Walker metric, 
\begin{equation}
    \label{eqn:CPLasTaylorina}
    w_\mathrm{CPL}(a) = w_0 + w_a(1-a)\,,
\end{equation}
with $w_0$ and $w_a$ constants. Formally, this expression can be viewed as the first-order truncation of a Taylor expansion of $w_{\rm DE}(a)$ around its present value, $a=1$. However, as emphasized by Linder (see, e.g., \rcite{Linder:2007wa,Linder:2024rdj}), the CPL form was originally motivated by its ability to capture and reproduce the dynamics of specific classes of dark energy models, namely thawing and freezing scalar-field models. Its interpretation as a simple Taylor expansion is therefore not always appropriate.

It is important to note, however, that within these classes of models the evolution of $w(a)$ occupies a restricted region of the $w$--$w'$ phase space. Outside this regime---for example, in models exhibiting phantom behavior or more complicated evolution---the CPL parametrization may not faithfully capture the underlying dynamics. This limitation was also emphasized in \rcite{Linder:2007wa,Linder:2024rdj}, where these regions of phase space were labeled as ``zones of avoidance.''

This implies that if the underlying dark energy model is a simple freezing or thawing single-scalar-field model, then the CPL parametrization provides a good description of the field dynamics, two parameters are sufficient, and the CPL should not be interpreted as a Taylor expansion. However, if the underlying model lies outside this region of parameter space (i.e., within the zones of avoidance), then the CPL parametrization does not accurately describe the underlying dynamics—even if it may formally provide a good fit to the data—and may therefore lead to misleading interpretations. In such cases, if the CPL parametrization is used, it should be regarded as a truncated Taylor expansion, bearing in mind that additional parameters may be required to capture the true dynamics of dark energy.

The CPL parametrization is often rewritten in terms of the redshift $z$,
\begin{equation}
\label{eqn:CPLofz}
w_\mathrm{DE}(z) = w_0 + w_a\frac{z}{1+z},
\end{equation}
where its form as a first-order Taylor expansion is less transparent. Observational collaborations typically report the values of $w_0$ and $w_a$, and include plots of the $68.3\%$ and $95.5\%$ confidence error ellipses in the $w_0$--$w_a$ plane. These plots are then widely used to infer whether the cosmological constant explains cosmic acceleration or whether some form of dynamical dark energy is at play.

As emphasized above, it is important to note that \eqref{eqn:CPLasTaylorina} is not a theory of dark energy. In particular, when one does not restrict attention to simple freezing and thawing single-scalar-field models, it should instead be regarded as a truncated two-dimensional phenomenological parametrization of $w_\mathrm{DE}(z)$. By projecting the data into this low-dimensional parameter space,
the CPL parametrization injects information into the inference that is not required by the data. In particular, by fixing all higher-order derivatives of $w(a)$ with respect to $a$ to zero, i.e.,
\begin{equation}
\label{eqn:CPLinjectedinfo}
\frac{\dd^p}{\dd a^p}w_\mathrm{CPL}(a) = 0,\quad \forall, p\in \mathbb{Z}^{\geq 2},
\end{equation}
the CPL parametrization effectively imposes strong priors on the functional form of $w_{\rm DE}(a)$, implicitly assuming that dark energy behaves as a freezing or thawing single scalar field. Information about the higher-order behavior of $w_{\rm DE}(a)$ is therefore supplied by the parametrization itself rather than by the observational data.

In principal, \eqref{eqn:CPLinjectedinfo} is rather an extraordinary amount of injected information, and so the question is what the effect would be of not assuming it.
There are alternative parametrizations/approaches that seek to address this problem, such as principal components \cite{Huterer:2002hy} and similar methods \cite{Bansal:2025ipo,Ormondroyd:2025exu,Berti:2025phi,Ormondroyd:2025iaf}, which represent our knowledge of $w_\mathrm{DE}(z)$ in terms of the independent pieces of information in the dataset; 
however, perhaps the simplest way to understand the effects of assuming  \eqref{eqn:CPLinjectedinfo} is to extend the CPL parametrization to higher orders in $1-a$, apply it to data, and then  marginalize over the Taylor-series coefficients of order two and higher.

For illustrative purposes, here we consider  extended CPL parametrizations up to third order in $1-a$:
\begin{equation}
    \label{eqn:CPL++}
    w_\mathrm{DE}(a) = w_0 + w_a(1-a) + w_b(1-a)^2 + w_c(1-a)^3\,.
\end{equation}
We then compare four parametrizations: 
(a) CPL$^-$: $w_0$ varying (i.e., $w_0$ is not necessarily $-1$ as in $\Lambda$CDM) but $w_a=w_b=w_c=0$;
(b)  CPL: $w_0$ and $w_a$ varying but $w_b=w_c=0$;
(c)  CPL$^+$: $w_0$, $w_a$, and $w_b$ varying but $w_c=0$, marginalized over $w_b$; 
and 
(d) CPL$^{++}$: $w_0$, $w_a$, $w_b$, and $w_c$ varying, marginalized over $w_b$ and $w_c$.

\para{Data and method.}
We use the Dark Energy Spectroscopic Instrument (DESI) DR1 baryon acoustic oscillations (BAO) data \cite{DESI:2024mwx}, the Pantheon+ compilation\footnote{We could have used the supernova data provided by the Dark Energy Survey (DES) collaboration, which supersede the Pantheon+ data. DES-SN5YR presented substantial new high-redshift supernova data and improved the techniques from Pantheon+, while DES-Dovekie further improved the cross-calibration between low- and high-redshift supernovae and made analysis improvements. However, as our paper primarily focuses on the limitations of the CPL parametrization it is not necessary to use the latest data set. Since all of the supernova data sets are consistent with each other, the results of this paper would not substantially change.} of 1550 type Ia supernovae, including all systematic and statistical uncertainties, along with a $z>0.01$ cut in order to minimize the effects of peculiar velocities \cite{Scolnic:2021amr}, and the {\it Planck} cosmic microwave background (CMB) data \cite{Planck:2018vyg} in the form of compressed distance measurements at recombination, also known as CMB shift parameters  (see \rcite{Zhai:2018vmm} for details). 
In order to be consistent with \rcite{DESI:2024mwx}, we also assume the approximate expression of the drag-epoch sound horizon of \rcite{Brieden:2022heh} and a big bang nucleosynthesis 
prior on the baryon density $\Omega_\mathrm{b}h^2=0.02218 \pm 0.00055$ \cite{Schoneberg:2024ifp}, where $\Omega_\mathrm{b}$ is the baryon density parameter and $h\equiv H_0/(100\,\mathrm{km}\,\mathrm{s}^{-1}\,\mathrm{Mpc}^{-1})$, with $H_0$ the present value of the Hubble expansion rate $H$. 
We have confirmed that this combination of data and likelihoods, along with our pipelines and codes, reproduce exactly the DESI constraints on the $\Lambda$CDM and CPL parameters provided in \rcite{DESI:2024mwx}.

We perform a Markov chain Monte Carlo (MCMC) analysis of CPL$^{-}$, CPL, CPL$^{+}$, and CPL$^{++}$ by allowing the matter density parameter $\Omega_\mathrm{m}$, the reduced Hubble constant $h$, and the $w_0$, $w_a$, $w_b$, and $w_c$ parameters (depending on the parametrization) to vary. With $\sim500$K MCMC points and after ensuring that the chains are well converged, we marginalize over $w_b$ for CPL$^{+}$ and over $w_b$ and $w_c$ for CPL$^{++}$ to obtain the two-dimensional $w_0$-$w_a$ $68.3\%$ and $95.5\%$ confidence contours, as well as the one-dimensional marginalized probability density functions for $w_0$ and $w_a$. 
From the MCMC chains we can also obtain the covariance matrix of all the parameters, which we then use along with standard error propagation to obtain the best-fit and $1\sigma$ error regions for $w_\mathrm{DE}(z)$, i.e., the equation-of-state parameter as a function of redshift. 

\begin{figure}[!t]
\includegraphics[width=0.5\textwidth]{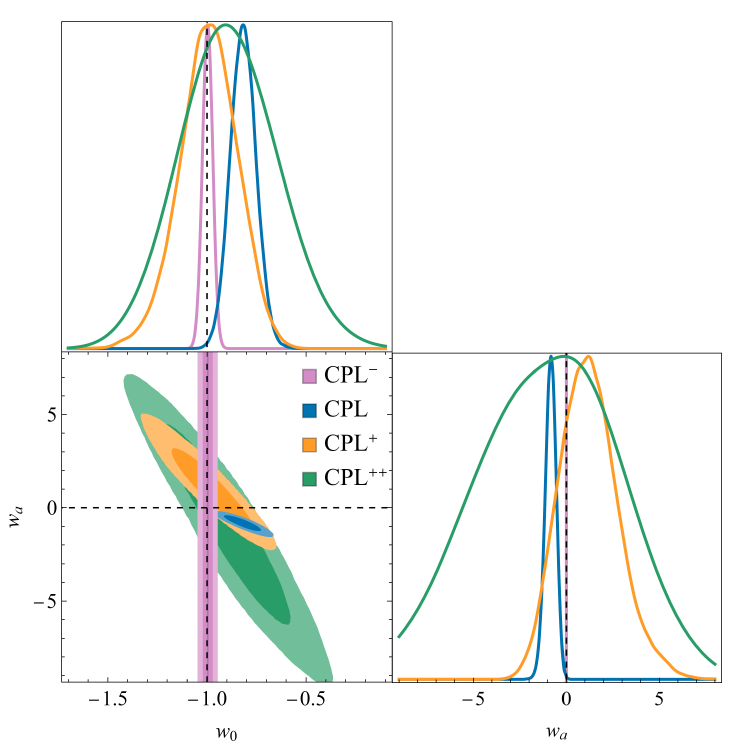}
\caption{A triangle plot showing the two-dimensional, marginalized $w_0$-$w_a$ $68.3\%$ and $95.5\%$ confidence error contours (lower left panel) and the one-dimensional, marginalized posterior probability distribution functions for $w_0$ (upper panel) and $w_a$ (lower right panel). The different colors correspond to the different cases considered in the present work: CPL$^{-}$ (magenta), where only $w_0$ is varied (i.e., $w_\mathrm{DE}$ is constant); CPL (blue), where only $w_0$ and $w_a$ are varied; CPL$^{+}$ (orange), where only $w_0$, $w_a$, and $w_b$ are varied; CPL$^{++}$ (green), where $w_0$, $w_a$, $w_b$, and $w_c$ are all varied.}
\label{fig:Fig1}
\end{figure}

\begin{figure}[!t]
\includegraphics[width=0.5\textwidth]{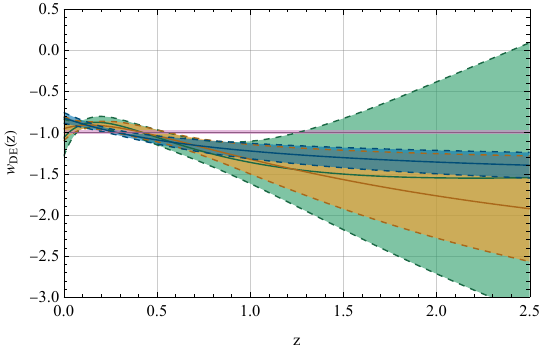}
\caption{The dark energy equation-of-state parameters $w_\mathrm{DE}(z)$ corresponding to the four parametrizations considered in the present work (CPL$^{-}$, CPL, CPL$^{+}$, and CPL$^{++}$) as a function of redshift $z$. The colors are as in Fig.~\ref{fig:Fig1}, the solid curves correspond to the best-fit values of the fitted parameters, and the color-shaded regions show the corresponding 1$\sigma$ error regions.}
\label{fig:Fig2}
\end{figure}

\para{Results}
We present in \cref{fig:Fig1} a traditional triangle plot showing the one-dimensional posterior probability density distributions for $w_0$ (upper panel) and $w_a$ (lower right panel), as well as the two-dimensional $w_0$-$w_a$ error ellipses at the $68.3\%$ and $95.5\%$ confidence levels (lower left panel) for our four parametrizations CPL$^{-}$, CPL, CPL$^{+}$, and CPL$^{++}$. \cref{fig:Fig2} displays the reconstructions of $w_\mathrm{DE}(z)$ that result from the four parametrizations; the solid curves are obtained using the best-fit values of the corresponding subset of parameters, while the colored regions are the $1\sigma$ error bands around the best-fit curves obtained by standard error propagation at each point:
\begin{equation}
    \sigma_{w_\mathrm{DE}(z)}^2 = \frac{\dd w_\mathrm{DE}}{\dd p_i}\,C_{ij}\, \frac{\dd w_\mathrm{DE}}{\dd p_j}\,,\quad \mathrm{for} \,\, p_i\in\{w_0,w_a,w_b, w_c\} \,,
\end{equation}
where the correlation matrix $C_{ij}$ is obtained directly from the MCMC chains after removing burn-in samples.

The contrast between the traditional CPL parametrization in blue and all three others in \cref{fig:Fig1} is striking.
In particular, it is clear that if one sets $w_b=w_c=0$, then there appears to be strong evidence against $(w_0,w_a)=(-1,0)$, i.e., the cosmological constant in the CPL parametrization, as the explanation of cosmic acceleration,
and in favor of a $w_\mathrm{DE}(z)$ with $w_0$ far from $-1$ but a $\dd w_\mathrm{DE}/\dd z<0$, leading to  phantom behavior ($w_\mathrm{DE}<-1$) at very moderate redshifts---see the blue band in \cref{fig:Fig2}.
However, if one allows even one additional term in the Taylor series (the CPL$^+$ parametrization), then the preferred value of $w_\mathrm{DE}$ today is consistent with $-1$, and the preferred slope of $w_\mathrm{DE}(z)$ is positive.  
In fact, if one marginalizes over the coefficient $w_b$ instead of setting it to 0, then the evidence against $(w_0,w_a)=(-1,0)$ falls below $2\sigma$ and if one allows two additional terms (the CPL$^{++}$ parametrization) and marginalizes over $w_b$ and $w_c$, then the evidence against  $(w_0,w_a)=(-1,0)$ falls below $1\sigma$.
In other words, the data is consistent with being dominated \textit{today} by constant vacuum energy. Additionally, \cref{fig:Fig2} shows that the evidence for phantom-like dark energy at high redshifts significantly weakens when one includes the higher-order terms required to probe the phantom region within the zones of avoidance.

This is not to say that there is no evidence against a pure cosmological constant. In \cref{tab:AICofCPLx}, we 
present the Akaike information criterion (AIC) \cite{Trotta:2008qt} for the CPL$^{-}$, CPL, CPL$^{+}$, and CPL$^{++}$ parametrizations, as well as for $\Lambda$CDM. We see that there is a definite decrease in AIC for the CPL parametrization over $\Lambda$CDM and CPL$^{-}$, suggesting that something in the data is arguing against $w_\mathrm{DE}(z)=-1$. 
On the other hand, CPL$^{+}$ has essentially the same AIC as CPL and leads to a radically different $w_\mathrm{DE}(z)$, especially at the range of redshifts where most of the information is---see Fig.~8 of \rcite{Lodha:2025qbg}. We 
also present in \cref{tab:AICofCPLx} the best-fit $\chi^2$ corresponding to $\Lambda$CDM, as well as to each of the four parametrized extensions. The $\chi^2$ for CPL$^{++}$ is only marginally smaller than for CPL$^{+}$, and consequently the AIC is somewhat higher, but the best-fit values of the phenomenological parameters  continue to shift and the shape of the reconstructed $w_\mathrm{DE}(z)$ changes still more.

It is perhaps appropriate at this point to note that all of the $\chi^2$ values in \cref{tab:AICofCPLx} are significantly lower than the number of degrees of freedom (dof) $\simeq 1600$ minus the number of parameters, so it is not actually clear that a parametrization with a lower $\chi^2$ is preferred, as the AIC would suggest. The reason for this is that in all the cases one finds $\chi^2/\mathrm{dof}<1$. Possibly this is due to overestimation of the supernovae errors; see \rcite{Nielsen:2015pga, Sah:2024csa}.

\begin{table}

    \centering
    \setlength\tabcolsep{10pt}
    \begin{tabular}{lcc}
    \hline
    Model/Parametrization~ & $\chi^2$ & AIC \\
    \hline
    $\Lambda$CDM & 1439.41 & 1443.41 \\
    CPL$^{-}$  & 1439.39 &  1445.39 \\
    CPL       & 1430.24 &  1438.24 \\
    CPL$^{+}$  & 1428.98 &  1438.98 \\
    CPL$^{++}$ & 1428.93 &  1440.93 \\
    \hline
    \end{tabular}
    \caption{$\chi^2$ values for the best-fit parameters of the standard $\Lambda$CDM model and its parametric extensions studied in the present work, as well as their corresponding AIC values. We count $\sim1600$ degrees of freedom.}\label{tab:AICofCPLx}
\end{table}

Returning to \cref{fig:Fig2},
we first note that as one moves to progressively higher-order parametrizations a visual inspection of the curves indicates that the error bands become wider. 
This is a well-known challenge with such reconstructions---as one adds more parameters, the bias in the reconstruction decreases but the uncertainty increases. It is a particular challenge for a function like $w_\mathrm{DE}(z)$, which is not measured directly at each $z$ but inferred by fitting data over a wide range of $z$.

Second, $w_\mathrm{DE}=-1$ is well within the $1\sigma$ error band for the constant $w$ parametrization (i.e., CPL$^{-}$), but appears to be outside the error band in certain ranges of $z$ for all three of CPL, CPL$^+$, and CPL$^{++}$. This is despite the fact that $w_\mathrm{DE}=-1$ is well within the $1\sigma$ error contours of the  CPL$^{++}$ parameter space. This apparent difference is an artifact of the reconstruction of the error bands for the dark energy equation-of-state parameter and, in fact, is a well-known problem of CPL-like parametrizations \cite{Alam:2004ip}. The reason for this apparent reduction in the errors at specific points is that as most $w_\mathrm{DE}(z)$ curves cross $w_\mathrm{DE}=-1$ at approximately the same redshift, this creates a \textit{sweet spot} \cite{Alam:2004ip}, as already noted in \rcite{Nesseris:2005ur} for the CPL parametrization.

\para{Conclusions}
By assuming a simple Taylor expansion for the dark energy equation-of-state parameter $w_\mathrm{DE}(z)$ and truncating it to first order, i.e., the CPL parametrization in regions of the equation-of-state phase space beyond simple freezing and thawing single-scalar-field models, there appears to be strong evidence against a vacuum energy that is currently constant. However, we have shown that, by allowing higher-order terms in the Taylor series and marginalizing over them, the evidence immediately falls below $1\sigma$.

We have similarly shown that allowing higher-order terms significantly weakens the evidence for phantom-like dark energy at high redshifts.
Since the CPL parametrization was not originally designed to describe phantom dark energy models---and such models indeed lie within the so-called zones of avoidance of the CPL parametrization as a physics-based model---any such behavior inferred from CPL fits should be interpreted with caution and may reflect limitations of the parametrization rather than properties of the underlying physics.

Our results immediately signal that projection effects are at play, and consequently, recent fits based on the two-parameter CPL Taylor expansion should be interpreted with great care, since truncating at first order introduces additional information that is not contained in the data, i.e., it implicitly assumes that the underlying dark energy model corresponds to a simple freezing or thawing single-scalar-field scenario. In other words, when the CPL parametrization is used in the zones of avoidance, one is effectively setting the higher-order parameters of the expansion (e.g., $w_b$, $w_c$, etc.) to zero by hand. This is equivalent to imposing a strong prior on these parameters.

The conclusion is clear: the CPL parametrization, by imposing that the second and higher derivatives of $w_\mathrm{DE}(a)$ with respect to $a$ vanish---information that is neither derived from the data nor necessarily required by physical models---unjustifiably excludes the possibility that the Universe is currently dominated by constant vacuum energy and also excludes many dynamical (non-phantom) dark energy models, such as minimally coupled quintessence (see, e.g., \rcite{Akrami:2025zlb}), as viable explanations for cosmic acceleration.
More specifically, it provides evidence for phantom-like dark energy at high redshifts, a class of models for which the CPL parametrization was not designed and should not be used, similarly to other regions of the dark energy zones of avoidance, for which additional parameters in the expansion are required to capture the underlying dynamics.

This is not an argument in favor of higher-order phenomenological parametrizations, but rather an argument that while phenomenological parametrizations may be useful for frequentist testing of models, they are dangerous when used to draw conclusions about how to repair models that fail such a test.
We also emphasize that we do not claim here that the DESI data do not imply the presence of  dynamical  dark energy---taking the data at face value, the Akaike information criterion suggests a preference for some evolution of the dark energy equation of state  at some time in the past (though the too-low $\chi^2/\mathrm{dof}$ should give one pause when drawing that conclusion). We instead stress that one should not use the two-parameter constraints on the current value of the equation of state, $w_0$, and its first derivative, $w_a$, based on the CPL Taylor expansion, to draw conclusions about theoretical models of dark energy, especially for the vast majority of models that lie beyond the regime of validity of the CPL parametrization as a physics-based model.

This calls for strong caution when the CPL parametrization is applied to current high-precision cosmological data. The CPL parametrization is still widely used as a figure of merit in analyses by major observational collaborations without clearly emphasizing these limitations; consequently, misunderstandings about its regime of validity can directly affect the physical interpretation of observational results.

It is also important to note that other concerns have been identified with the use of the standard BAO analysis techniques, especially in the context of extensions to standard flat-$\Lambda$CDM cosmology (cf.  \rcite{Anselmi:2018vjz,Anselmi:2022exn}).

Moreover, one should be careful when using similar two-parameter extensions to test $\Lambda$CDM and to draw conclusions about its dark energy alternatives. Examples are some of the parametrizations studied in \rcite{Lodha:2025qbg} (referred to as BA, EXP, and LOG in Table II of that work), which are all effectively low-redshift parametrizations that behave almost exactly like the CPL parametrization at low redshifts, implying that they are also prone to the same issues discussed in the present work.

Data-derived alternatives, such as principal components, can be used to test the cosmological constant $\Lambda$ as the explanation of cosmic acceleration in a frequentist way, while physical models can be compared to $\Lambda$ in a Bayesian way, but one should not confuse low-dimensional phenomenological parametrizations for physical models.

{\bf Note added:} While this manuscript was in preparation, new (DR2) cosmology results of DESI \cite{DESI:2025zgx,Lodha:2025qbg} appeared. We did not update our analysis with the DR2 data since the DR1 data were sufficient for the purposes of the present work.

\parait{Code availability} All the numerical codes and MCMC chains used in this analysis will be made publicly available at \url{https://github.com/snesseris/DESI_CPL} upon the publication of the paper.

\begin{acknowledgments}
We thank G.~Alestas and S.~Anselmi for helpful discussions. S.N. acknowledges support from the research project PID2021-123012NB-C43 and the Spanish Research Agency (Agencia Estatal de Investigaci\'on) through the grant IFT Centro de Excelencia Severo Ochoa Grant No. CEX2020-001007-S, funded by MCIN/AEI/10.13039/501100011033.
Y.A. acknowledges support by the Spanish Research Agency (Agencia Estatal de Investigaci\'on)'s grant RYC2020-030193-I/AEI/10.13039/501100011033, by the European Social Fund (Fondo Social Europeo) through the  Ram\'{o}n y Cajal program within the State Plan for Scientific and Technical Research and Innovation (Plan Estatal de Investigaci\'on Cient\'ifica y T\'ecnica y de Innovaci\'on) 2017-2020, by the Spanish Research Agency through the grant IFT Centro de Excelencia Severo Ochoa Grant No. CEX2020-001007-S funded by MCIN/AEI/10.13039/501100011033, and by the Spanish National Research Council (CSIC) through the Talent Attraction grant 20225AT025. 
G.D.S.\ is supported by grant DESC0009946 from the US DOE, and thanks the IFT for its frequent and generous hospitality, during which conversations leading to this paper occurred. This work made use of the IFT Hydra cluster. 
\end{acknowledgments}

\bibliography{bibliography}
\end{document}